\documentclass[oribibl]{llncs} \pdfoutput=1 \usepackage{graphicx}
 \usepackage{amssymb}

\begin{document}

\pagestyle{plain}

\makeatletter
\def\imod#1{\allowbreak\mkern10mu({\operator@font mod}\,\,#1)}
\makeatother

\newcommand{\eprint}[1]{{\tt #1}} \newcommand{\eqn}[1]{(\ref{eq.#1})}
\newcommand{\bra}[1]{\mbox{$\left\langle #1 \right|$}}
\newcommand{\ket}[1]{\mbox{$\left| #1 \right\rangle$}}
\newcommand{\braket}[2]{\mbox{$\left\langle #1 | #2 \right\rangle$}}
\newcommand{\av}[1]{\mbox{$\left| #1 \right|$}}
\newcommand{\ve}{\varepsilon} \newcommand{\osc}{{\mbox{\rm \scriptsize
      osc}}} \newcommand{\tot}{{\mbox{\rm \scriptsize tot}}}
\newcommand{\lga}{{\mbox{\rm \scriptsize LGA}}}
\newcommand{\swap}{{\mbox{\rm \scriptsize swap}}}
\newcommand{\ground}{{\mbox{\rm \scriptsize ground}}}
\newcommand{\cycle}{{\mbox{\rm \scriptsize cycle}}}
\newcommand{\particle}{{\mbox{\rm \scriptsize particle}}}
\newcommand{\internal}{{\mbox{\rm \scriptsize internal}}}
\newcommand{\nonrel}{{\mbox{\rm \scriptsize non-rel}}}
\newcommand{\twoparticles}{{\mbox{\rm \scriptsize 2p}}}
\newcommand{\block}{{\mbox{\rm \scriptsize block}}}
\newcommand{\blockchange}{{\mbox{\rm \scriptsize block-change}}}
\newcommand{\statechange}{{\mbox{\rm \scriptsize state-change}}}
\newcommand{\even}{{\mbox{\rm \scriptsize even}}}
\newcommand{\odd}{{\mbox{\rm \scriptsize odd}}}
\newcommand{\rt}{{\mbox{\rm \scriptsize R}}}
\newcommand{\lt}{{\mbox{\rm \scriptsize L}}}
\newcommand{\ball}{{\mbox{\rm \scriptsize ball}}}
\newcommand{\shift}{{\mbox{\rm \scriptsize shift}}}
\newcommand{\D}[2]{\frac{\partial #2}{\partial #1}}

\newcommand{\sinc}{{\mbox{\rm sinc}}\:}
\newcommand{\sincs}{{\mbox{\rm sinc$^2$}}}

\renewcommand{\max}{{\mbox{\rm \scriptsize max}}}
\renewcommand{\min}{{\mbox{\rm \scriptsize min}}}

 \newcommand{\figt}[3]{
 \begin{figure}[tb]
  $$#2$$\relax
  \caption{#3}
 \label{fig.#1}
 \end{figure}                 }

\setlength{\fboxsep}{.1pt} \setlength{\fboxrule}{.1pt}


\title{The Ideal Energy of Classical Lattice Dynamics}


 \author{Norman Margolus}

\institute{Massachusetts Institute of Technology, Cambridge MA, USA.
  \email{nhm@mit.edu}}

\date{}

\maketitle

 \begin{abstract} 

{\em We define, as local quantities, the least energy and momentum
  allowed by quantum mechanics and special relativity for physical
  realizations of some classical lattice dynamics.  These definitions
  depend on local rates of finite-state change.  In two example
  dynamics, we see that these rates evolve like classical mechanical
  energy and momentum.}

\end{abstract}


\section{Introduction}\label{sec.intro}

Despite appearances to the contrary, we live in a finite-resolution
world. A finite-sized physical system with finite energy has only a
finite amount of distinct detail, and this detail changes at only a
finite rate~\cite{bekenstein,max-speed,max-avg}.  Conversely, given a
physical system's finite rates of distinct change in time and space,
general principles of quantum mechanics define its minimum possible
average energy and momentum.  We apply these definitions to classical
finite-state lattice dynamics.

\subsection{Ideal Energy}

It was finiteness of distinct state, first observed in thermodynamic
systems, that necessitated the introduction of Planck's constant $h$
into physics \cite{planck}.  Quantum mechanics manages to express this
finiteness using the same continuous coordinates that are natural to
the macroscopic world.  Describing reality as superpositions of waves
in space and time, finite momentum and energy correspond to
effectively finite bandwidth; hence finite distinctness.  For
example~\cite{max-avg}, the average rate $\nu$ at which an isolated
physical system can traverse a long sequence of distinct states is
bounded by the average (classical) energy $E$:
\begin{equation}\label{eq.nu}
\nu \le 2\,E/h\,,
\end{equation}
taking the minimum possible energy to be zero.  Here $E/h$ is the
average frequency of the state, which defines a half-width for the
energy frequency distribution.  If we compare \eqn{nu} in two frames,
we can bound the average rate $\mu$ of changes not visible in the rest
frame, and hence attributable to overall motion:
\begin{equation}\label{eq.mu}
\mu \le 2\,pv/h\,.
\end{equation}
Here $p$ is the magnitude of a system's average (classical) momentum,
which is also a half-width for a (spatial) frequency distribution; $v$
is the system's speed.

These kinds of constraints are sometimes referred to as uncertainty
bounds, but they in no way preclude precise finite-state evolution.
Given rates of change, these bounds define ideal (minimum achievable)
average energy and momentum for finite state systems, emulated as
efficiently as possible (with no wasted motion or state) by
perfectly-tailored quantum hamiltonians \cite{max-avg,emulation}.

Clearly there can never be more {\em overall spatial change}~$\mu$
than {\em total change} $\nu$ in a physical evolution: this is
reflected in $pv/E = (v/c)^2$.  From this and \eqn{mu},
\begin{equation}\label{eq.erel}
E\ge (h\mu/2)/(v/c)^2\,.
\end{equation}
Thus for a given rate $\mu$ of overall motional change, $E$ can only
attain its minimum possible value if the motion is at the speed of
light; then no energy is invested in rest-frame dynamics (rest
energy).  In a finite-state dynamics with several geometrically
related signal speeds, to minimize all energies \eqn{erel} the fastest
signals must move at the speed of light. If we then want to realize
the dynamics running faster, we must put the pieces of the system
closer together: we can increase $p$ in \eqn{mu}, but not $v$.  Of
course in finite-state models of particular physical systems,
realistic constraints on speeds and separations may require higher
energies.

These bounds can be used to define ideal local energies and momenta
for some invertible lattice dynamics, determined by rates of
distinct change.

\subsection{Local Change}

We restrict our attention to finite-state lattice dynamics that
emulate the locality, uniformity and microscopic invertibility of
physical law: invertible cellular automata (CA).  We assume the
dynamics is defined as a regular arrangement of invertible
interactions (logic gates), repeated in space and time, each of which
independently transforms a localized set of state variables.

This kind of CA format, where the state variables are always updated
in independent groups, has sometimes been called partitioning CA, and
encompasses a variety of lattice formats that have been used to model
physical dynamics
\cite{phys-like,cc,cam-book,chopard-book,rothman-book,rivet-book,salt,nks}.
It is interesting that all globally invertible CA can be recast in
this physically realistic format, as a composition of independent
invertible interactions, even if the CA was originally defined as a
composition of non-invertible operations on overlapping neighborhoods
\cite{ica,kari,lose}.  Historically, CA originated as physics-like
dynamics {\em without} invertibility \cite{ulam,von-neumann,zuse}.

Now, in the energy bounds above, only rates of change matter, not the
amount of state updated in a single operation.  This is unrealistic.
We can define a large-scale synchronous dynamics, where the global
rate of state change is independent of the size of the system.
Physically, total energy must be bounded by the total rate of local
changes, since each independent local update also obeys an energy
bound.  We resolve this conflict by allowing synchronous definition,
but counting the global average rate of distinct change as if local
updates were non-synchronous---which would in fact be true in most
relativistic frames.

There is also an issue of what not to count.  For a dynamics defined
by a set of gate operations, it might seem natural to include, in the
minimum, energy required to construct the gates and to turn them on
and off.  This is the energy needed to construct a perfectly-tailored
hamiltonian.  Here we ignore this construction energy, and discuss the
ideal case where the hamiltonian is given for free (as part of
nature), and we only need to account for energy required by state
change within the dynamics.

\subsection{Two Examples}



In the remainder of this paper, we introduce and discuss two
2$\times$2 block partitioning CA ({\em cf.} \cite{bcq}).  These
dynamics are isomorphic to classical mechanical systems, and are
simple enough that it is easy to compare energetic quantities, defined
by local rates of state change, with classical ones.

The first example is a scalable CA version of the Soft Sphere Model
\cite{ssm}, which is similar to Fredkin's classical mechanical
Billiard Ball Model \cite{bbm}.  This digital system emulates the
integer time behavior of an idealized classical mechanical system of
elastically colliding balls, and is computation universal.  The CA is
scalable in that square blocks of ones (balls) of any size can be
collided to simulate a billiard ball computation.  This model has not
been published before.

The second example is a CA model of an elastic string that exhibits
simple-harmonic motion and exactly emulates the continuum wave
equation at integer times, averaged over pairs of adjacent sites.
This model has been discussed
before~\cite{cc,hh,toffoli-action,marg-thesis}, but the analysis of
overall translational motion, ideal energy, and their relativistic
interpretation, have not been previously published.

\section{Scalable Soft Sphere CA}

Many CA dynamics can be interpreted as the integer-time behavior of a
continuous classical mechanical system, started from an exactly
specified initial state.  This is true, for example, for lattice gas
models of fluids.  Such {\em stroboscopic} classical mechanical CA
inherit, from their continuous counterparts, conserved quantities such
as energy and momentum that we can compare to ideal quantities
determined by local rates of state change.  Of course the continuum
models we have in mind would be numerically unstable if actually run
as continuous dynamics, but this issue is not inherited by the
finite-state CA \cite{toffoli-pde}.

A famous stroboscopic dynamics of this sort is Fredkin's billiard ball
model of computation, in which hard spheres moving in a plane, each
with four possible initial velocities, are restricted to a square
lattice of initial positions.  At each integer time, the system is
again in such a configuration. To guarantee this property without
additional restrictions on initial states, we let billiard balls pass
through each other in some kinds of collisions, without interacting.

Figure~\ref{fig.ssm} shows a variant of this model in which the balls
are much more compressible, so collisions deflect paths inward rather
than outward.  This variant has the advantage that it is more directly
related to a simple partitioning CA ({\em cf.} \cite{phys-like}).  In
the collision illustrated in Fig.~\ref{fig.ssm}a balls enter from the
left with a horizontal component of velocity of one column per time
unit, so consecutive moments of the history of a collision occur in
consecutive columns.

\figt{ssm}{
\begin{array}{c@{\hspace{.4in}}c@{\hspace{.4in}}c@{\hspace{.4in}}c}
\includegraphics[height=1.3in]{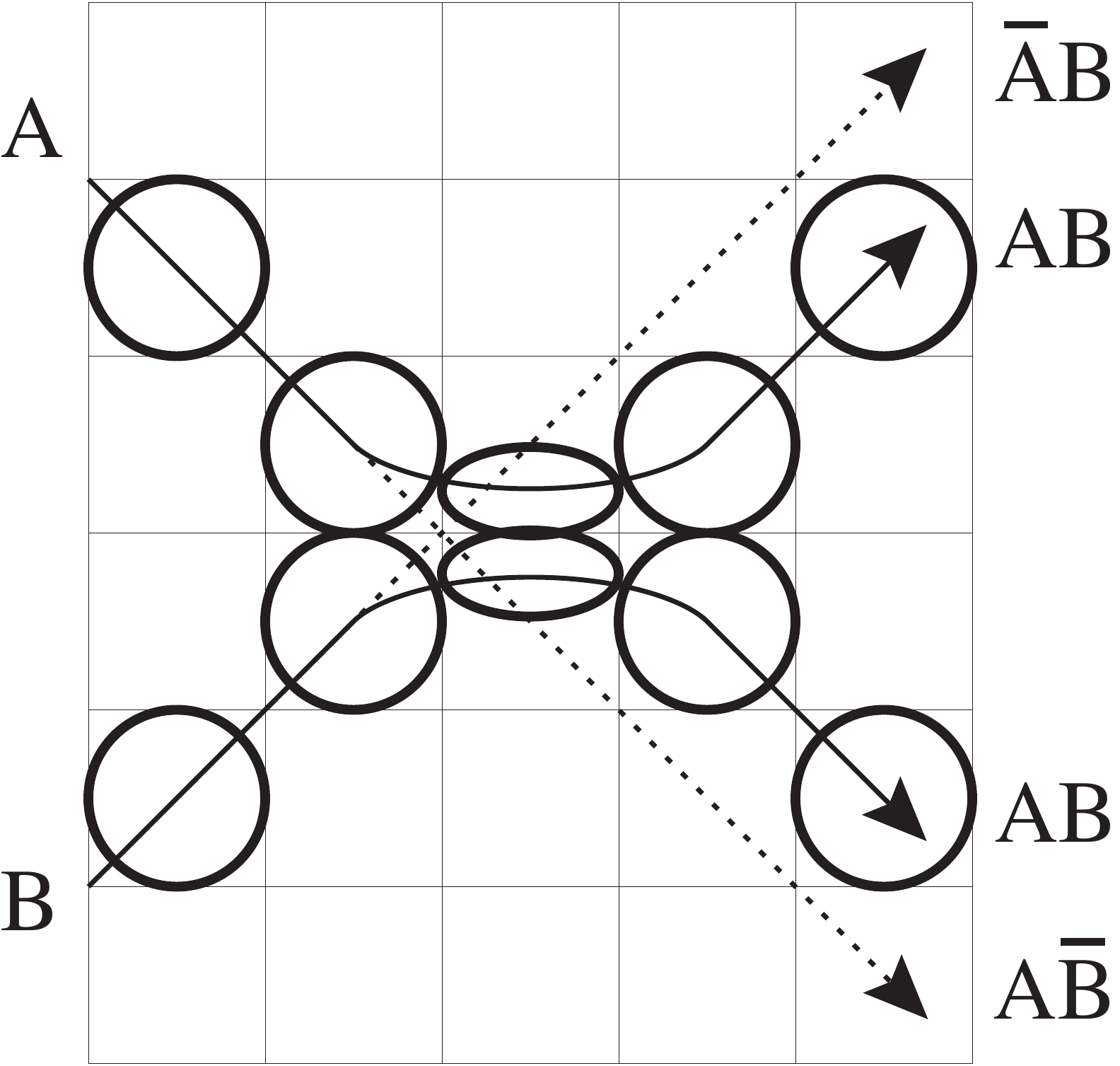} &
\includegraphics[height=1.3in]{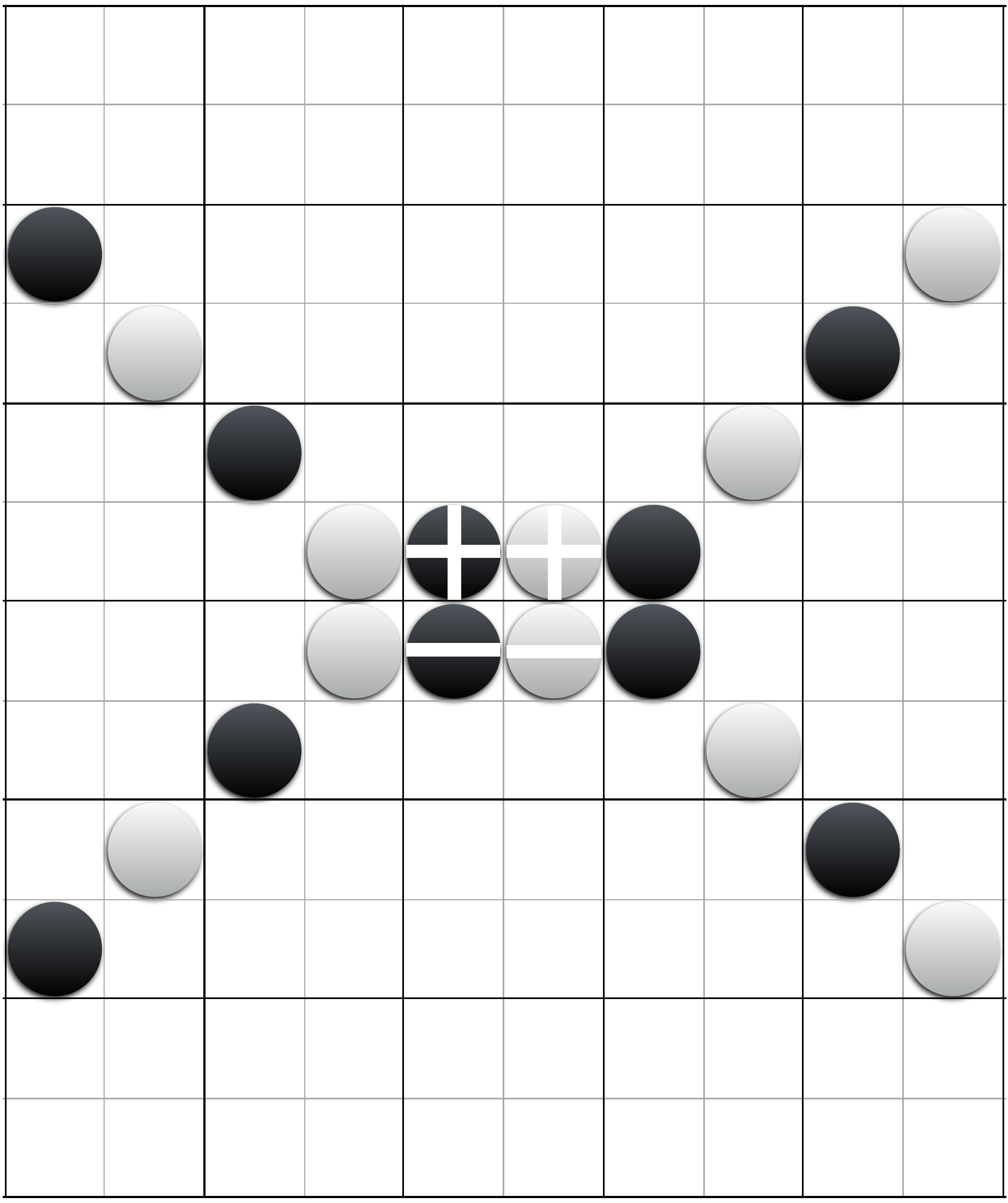} &
\includegraphics[height=1.3in]{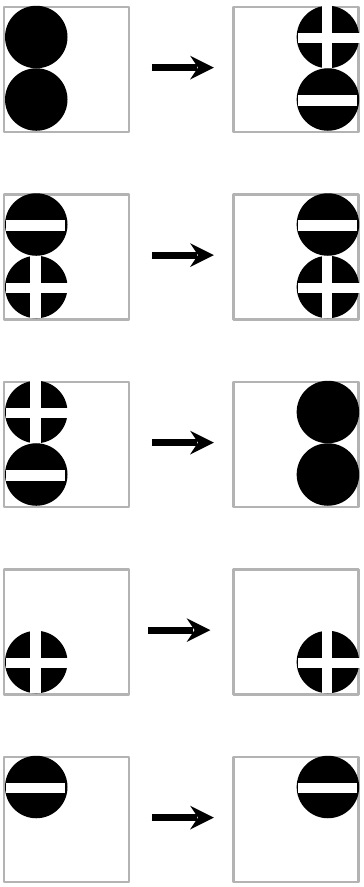} &
\includegraphics[height=1.3in]{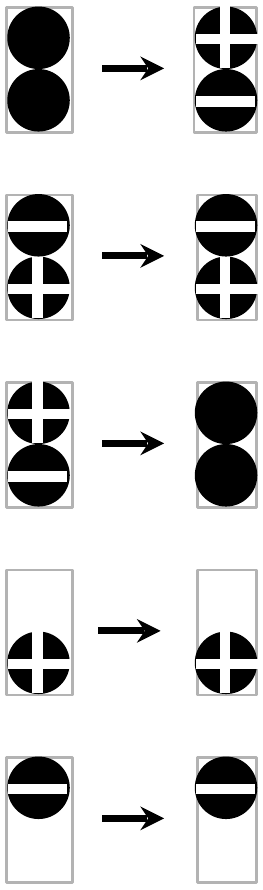} \\[.1in] \mbox{\bf (a)} &
\mbox{\bf (b)} & \mbox{\bf (c)} & \mbox{\bf (d)} \\
\end{array}}
{Scalable Soft Sphere dynamics.  (a) Stroboscopic view of continuous
  classical collision, one time step per column.  (b) Finite state
  collision, with particles drawn lighter at odd time-steps.  (c) 2D
  partitioning rule. Only these cases (and their rotations)
  interact. Otherwise, all particles move diagonally, unchanged. (d)
  1D version of the rule.}

The collision shown is energy and momentum conserving, and compression
and rebound take exactly the time needed to displace the colliding
balls from their original paths onto the paths labeled $AB$.  If a
ball had come in only at $A$ with no ball at $B$, it would have left
along the path labeled $A\bar{B}$: the collision acts as a universal
logic gate.

Figure~\ref{fig.ssm}b shows a realization of the collision as a simple
partitioning dynamics.  Each time step in (a) corresponds to two in
(b), and again particles are shown at each integer time---drawn dark
at even times and light at odd. The rule (c) is inferred from (b),
interpreting that diagram as showing the positions of two streams of
colliding particles at one moment (dark), and their positions at the
next moment (light).  All particles move diagonally across a block,
unchanged, in the cases not defined explicitly in (c).  If this rule
is applied to just the dark particles in each of the dark-bordered
2$\times$2 blocks in (b), ignoring the light particles, it moves them
to the light positions; applied to just the light particles in each of
the light-bordered blocks, it moves them to the dark positions.  The
dynamics alternately applies the rule to these two partitions.  To
also allow collisions like (b) for streams of balls arriving from the
right, top, or bottom, we define the rule (c) to have discrete
rotational symmetry: in each of the cases shown in (c), each of the
four $90^\circ$ rotations of the pattern on the left turns into the
corresponding rotation of the pattern on the right.

Note that (b) can also be interpreted as showing a time history of a
collision of two particles in a one-dimensional partitioning dynamics
(the center of mass dynamics). Then we get the rule (d), with the
cases not explicitly shown interchanging the two cell values.  Three
dimensional versions of the dynamics can be constructed as in
\cite{ssm}.

It is not surprising that a time-independent continuous dynamics turns
into a time-dependent discrete partitioning.  In the continuous model,
balls approach a locus of possible collision, interact independently
of the rest of the system, and then move away toward a new set of
loci.  The partitions in the continuous case are just imaginary boxes
we can draw around places within which what happens next doesn't
depend on anything outside, for some period of time.  Thus it is also
not surprising that we can assign a conserved energy to partitioning
dynamics.


\subsection{Ideal Energy and Momentum}

For a physical realization of the SSS dynamics, let $\tau$ be the time
needed for gate operations to update all blocks of one partition, and
let $v_0$ be the average speed at which the physical representation of
a fastest-moving particle travels within the physical realization of
the CA lattice (assuming discrete isotropy).


Equations \eqn{mu} and \eqn{erel} define an ideal (minimum) momentum
and energy for a block in which there is a distinct overall spatial
change and direction of motion. Clearly these ideal quantities are
conserved overall in collisions, since freely moving particles move
diagonally at $v_0$ before and after.  Are they also conserved in
detail during collisions?

When two freely moving particles enter a single block in the collision
of Fig.~\ref{fig.ssm}b, the number of block changes is reduced: one
instead of two.  The ideal magnitude of momentum for each freely
moving particle before the collision is $p_1=(h/2\tau)/v_0$.  For two
colliding particles moving horizontally within a block the ideal is
$p_2=(h/2\tau)/(v_0/\sqrt{2})=2 p_1/\sqrt{2}$, which is the same as
the net horizontal momentum before the collision.  Ideal energy is
similarly conserved.

Note, however, that the separate horizontal motions of the $+$ and $-$
particles during the next step of the collision of Fig.~\ref{fig.ssm}b
imply an increase in the minimum energy and momentum for that step.
This effect becomes negligible as we enlarge the scale of the objects
colliding.

\subsection{Rescaling the Collision}

If two columns of $k$ particles are collided in the SSS dynamics, then
the resulting collision just shifts the output paths by $k$ positions
along the axis of the collision.  This is illustrated in
Fig.~\ref{fig.block-coll}a for $k=3$.  Thus $k\times k$ blocks of
particles collide exactly as in the classical collision of
Fig.~\ref{fig.ssm}a: the SSS CA can perform logic with
diagonally-moving square ``balls'' of any size.  When two balls of
equal size meet ``squarely,'' moving together along a horizontal axis,
each pair of columns evolves independently of the rest; colliding
along a vertical axis, pairs of rows evolve independently.  Square
balls can participate in both kinds of collisions.

\figt{block-coll}{%
\begin{array}{c@{\hspace{.19in}}c@{\hspace{.19in}}c}
\includegraphics[height=1.09in]{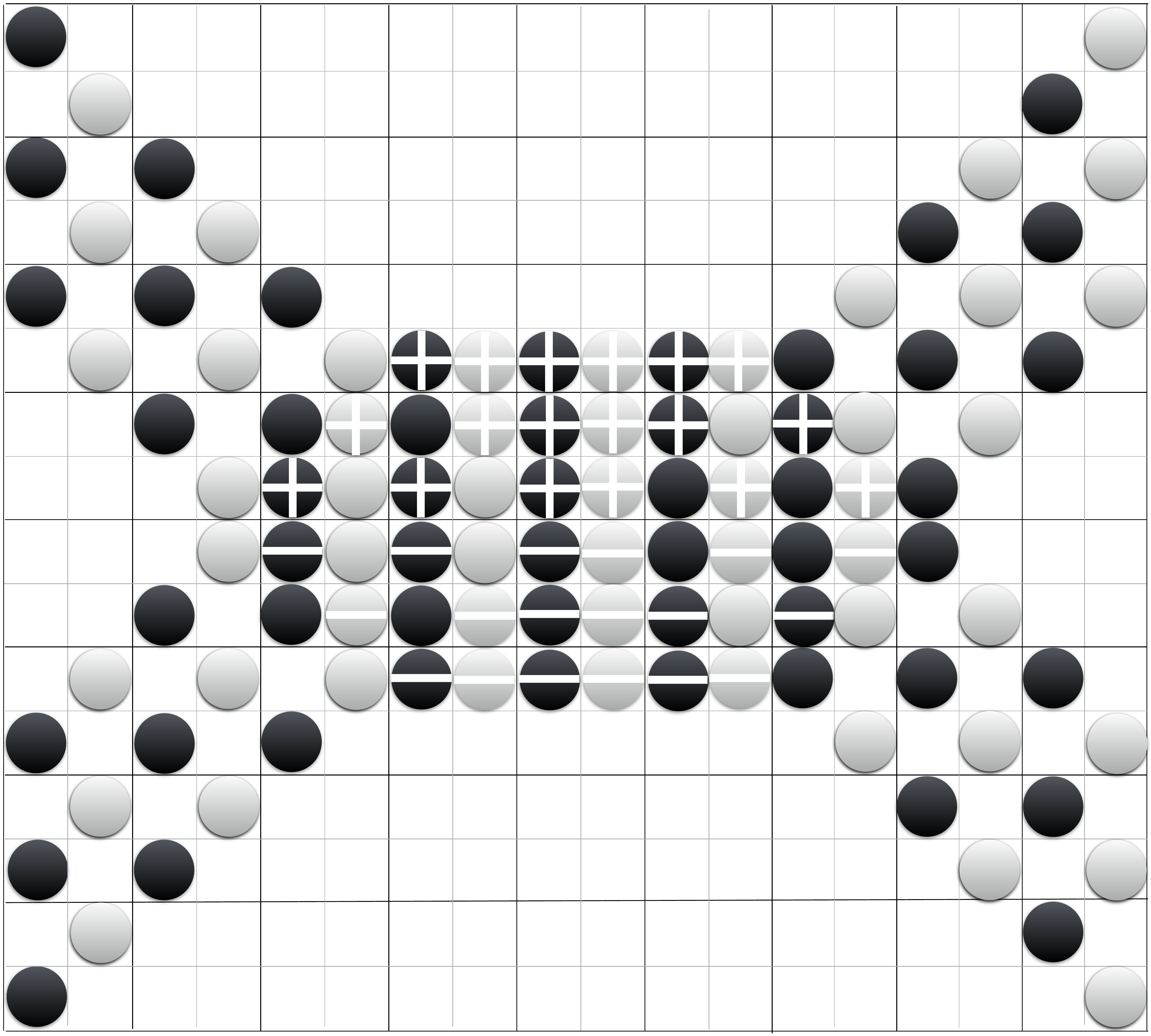} &
\includegraphics[height=1.09in]{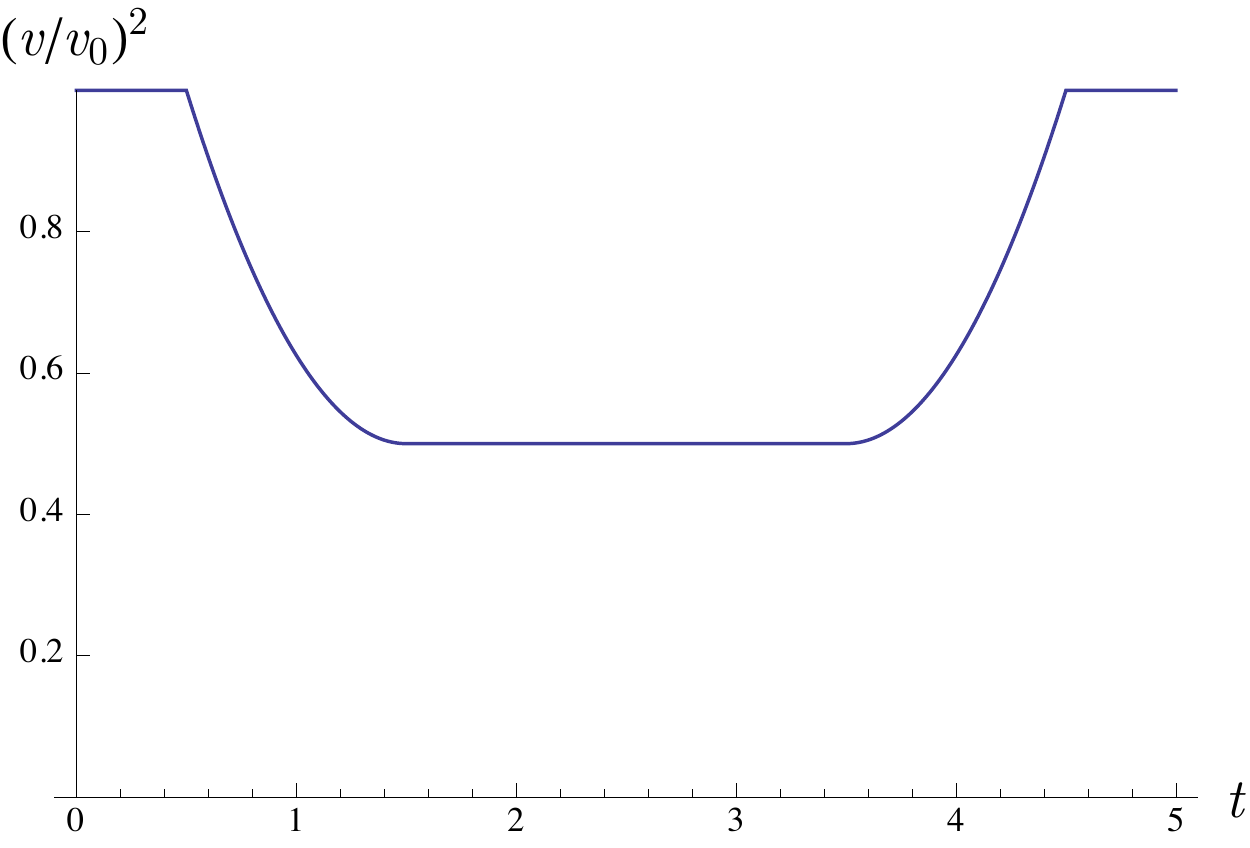} &
\includegraphics[height=1.09in]{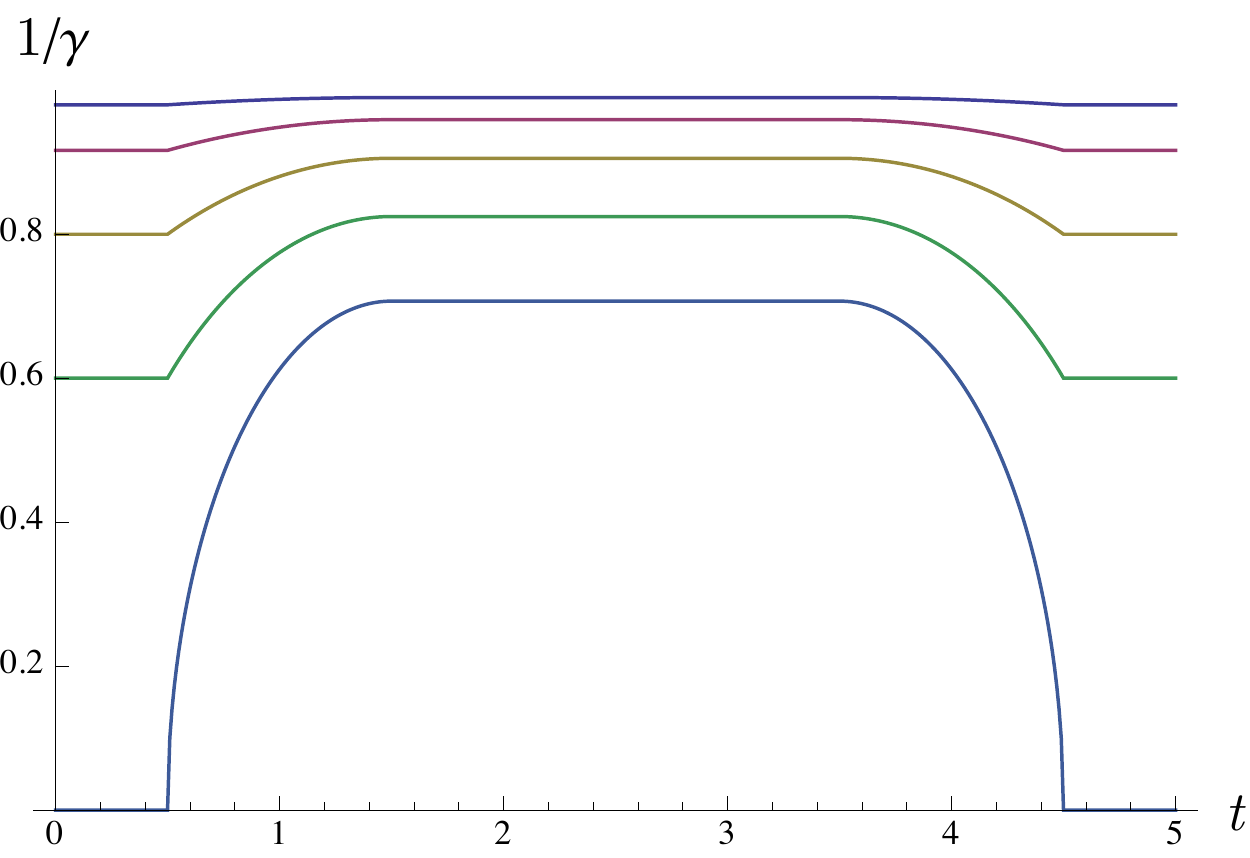} \\ \mbox{\bf (a)} &
\mbox{\bf (b)} & \mbox{\bf (c)} \\
\end{array}}
{Multiple collisions in the SSS dynamics. (a) Colliding columns of
  particles are displaced horizontally by the height of the column.
  (b) Each column is slowed down by the collision.  (c) The fraction
  of energy that is mass during a collision decreases with increasing
  initial particle speed $v_0$ (from top, 20\% of $c$, 40\%, 60\%,
  80\%, 100\%).}

During such a collision, from the blocks that change we can infer a
net momentum and hence a velocity for the motion of each colliding
ball: Fig.~\ref{fig.block-coll}b illustrates this for $k=100$, with
the time unit being the time for a freely moving $k\times k$ ball to
travel its length (and width).  Looking at just the changes in the top
half of (a), we determine the magnitude and direction of minimum
average momentum for each block that changes using \eqn{mu}, and hence
determine a total momentum.  Half of the conserved total energy is
associated with each ball, so $v/v_0 = v E_\ball /v_0 E_\ball =p/p_0$
gives the magnitude of velocity of a ball as a function of time, as
number and type of changes evolve.  This is plotted in (b).

The fraction $1/\gamma$ of the total energy $E$ that is mass energy
depends on $(v/c)^2 = (v/v_0)^2(v_0/c)^2$.  Thus given (b), it depends
on an assumption about the value of $v_0/c$.  The fraction $1/\gamma$,
as a function of time in the $k=100$ collision, is plotted in
Fig.~\ref{fig.block-coll}c under different assumptions.  The bottom
case, $v_0=c$, has the greatest range but the smallest value at all
times.  The top case is $v_0=.2c$.  As expected, the faster the speed
of the fastest signals, the less the energy tied up in mass, hence the
smaller the total energy.  Ideally, $v_0=c$.

\section{Elastic String CA}\label{sec.waves}

In this second example we discuss a classical finite state model of
wave motion in an elastic string.  This stroboscopic classical
mechanical model exactly reproduces the behavior of the
time-independent one-dimensional wave equation sampled at
integer-times and locations.  As in the SSS example, a continuous
model is turned into a finite state one by restricting the initial
state (in this case the initial wave shape) to a perfect discrete set
of possible initial configurations, and this constraint reappears at
each integer time.  In the continuum limit the discrete constraint on
the wave shape disappears; the exactness of the wave dynamics itself
(at discrete times) is independent of this limit.

The elastic string CA uses partitioning, but in a different way than
the SSS CA: here the partitioning actually constrains the continuous
classical dynamics used to define the CA, but in a way that never
affects the classical energy.  In the SSS case, the time dependence
associated with the partitioning completely disappears in the
continuous classical-mechanical version of the dynamics.

\subsection{Discrete Wave Model}

Consider an ideal continuous string for which transverse displacements
exactly obey the wave equation.  In Figure~\ref{fig.waves}a we show an
initial configuration with the string stretched between equally spaced
vertical bars.  The set of initial configurations we're allowing are
periodic, so the two endpoints must be at the same
height.\footnote{Unless the right and left edges of the space itself
  are joined with a vertical offset.}  Any configuration is allowed as
long as each segment running between vertical bars is straight and
lies at an angle of $\pm 45^\circ$ to the horizontal.

\figt{waves}{%
\begin{array}{c@{\hspace{.3in}}c@{\hspace{.3in}}c}
\includegraphics[width=1.4in]{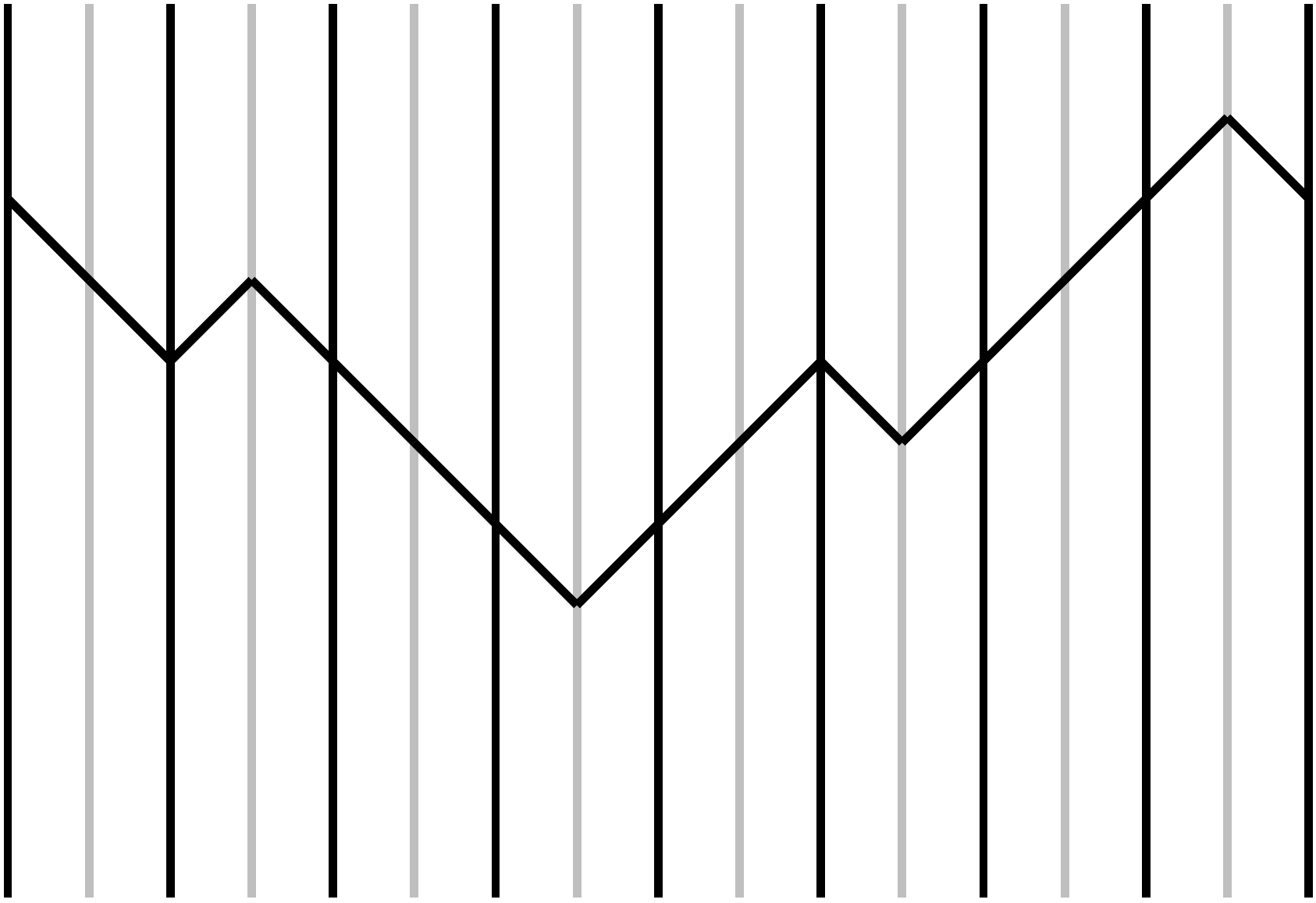} &
\includegraphics[width=1.4in]{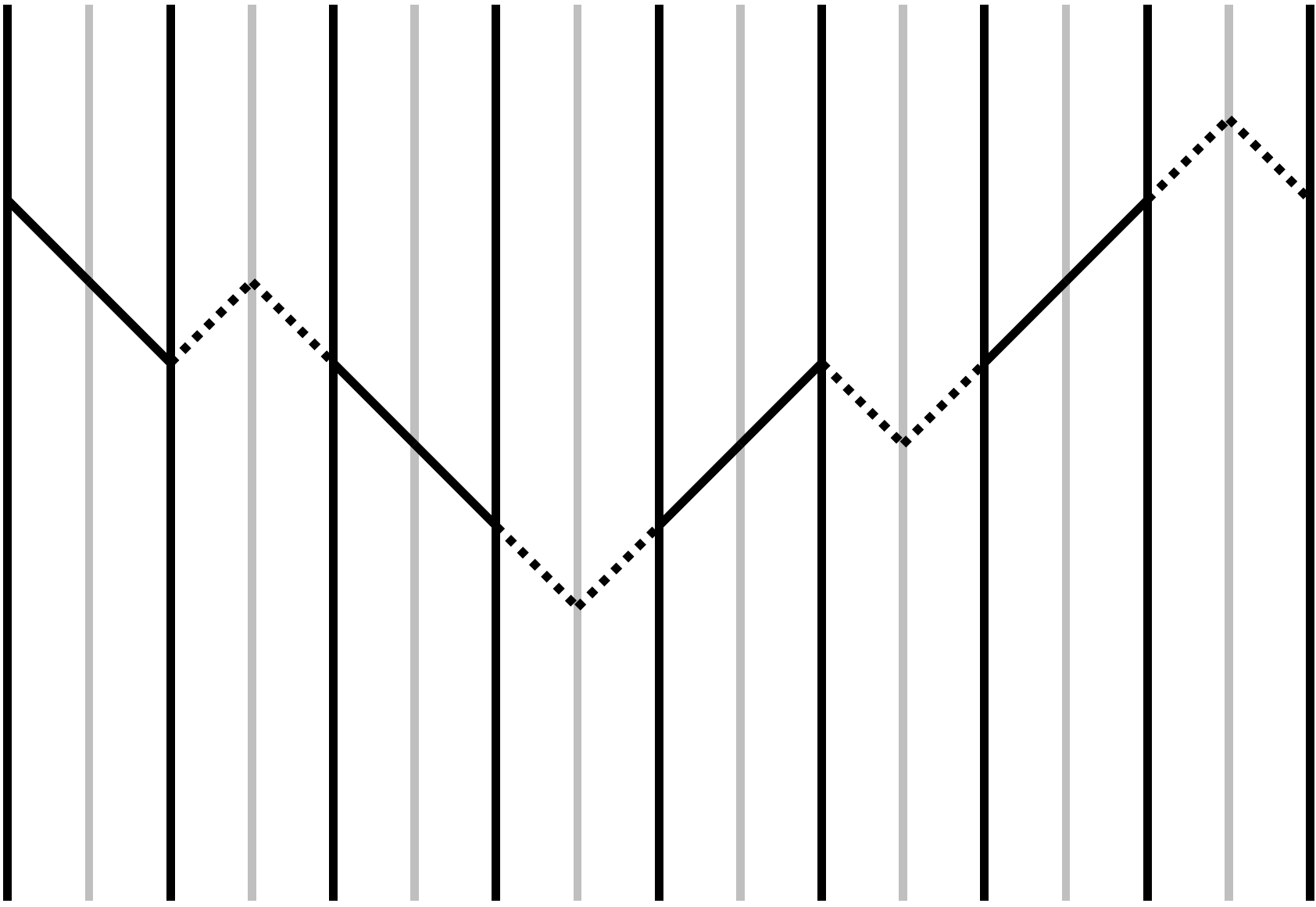} &
\includegraphics[width=1.4in]{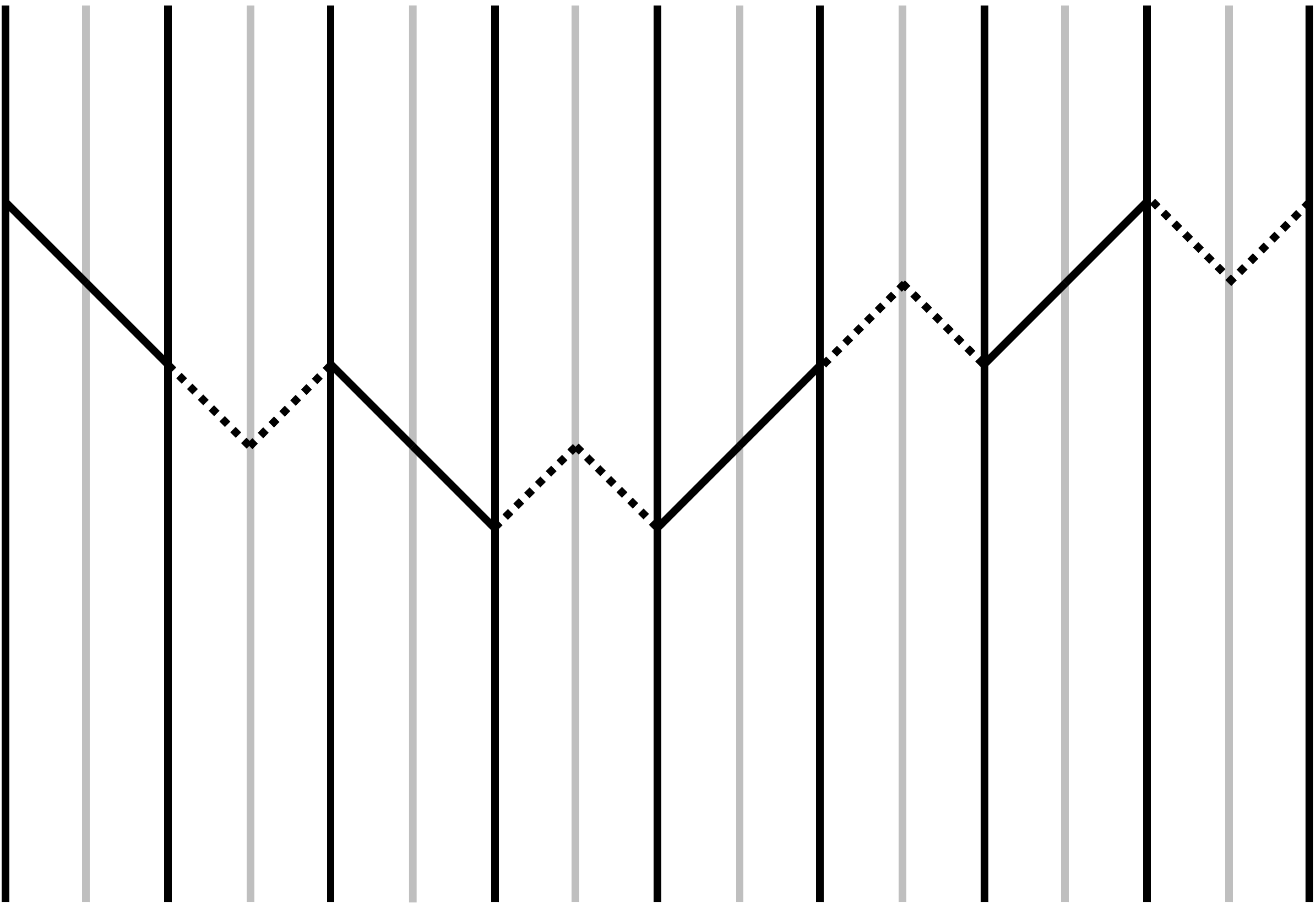} \\ \mbox{\bf (a)} & \mbox{\bf
  (b)} & \mbox{\bf (c)} \\
\end{array}}
{Discrete wave dynamics. Elastic string is held fixed where it crosses
  black bars.}

Initially the string is attached at a fixed position wherever it
crosses a vertical bar.  We start the dynamics by releasing the
attachment constraint at all of the gray bars.  The attachment to the
black bars remains fixed.  In Figure~\ref{fig.waves}b the segments
that are about to move are shown with dotted lines: the straight
segments have no tendency to move.  Under continuum wave dynamics, the
dotted segments all invert after some time interval $\tau$.  This will
be our unit of time for the discrete dynamics.  The new configuration
at the end of this interval is shown in Figure~\ref{fig.waves}c, with
segments that have just moved dotted.  At this instant in time all
points of the string are again at rest and we are again in an allowed
initial configuration.  Now we interchange the roles of the black and
gray bars and allow the segments between adjacent gray bars to move
for a time interval $\tau$.  The dynamics proceeds like this,
interchanging the roles of the black and gray bars after each interval
of length $\tau$.  Since attachments are always changed at instants
when all energy is potential and the string is not moving, the
explicit time dependence of the system doesn't affect classical energy
conservation.

We express this dynamics as a purely digital rule in
Figure~\ref{fig.waves-rule}.  In Figure~\ref{fig.waves-rule}a we show
a wave with the black bars marking the attachments for the next step.
To simplify the figure we have suppressed the gray bars---they are
always situated midway between the black bars and so don't need to be
shown.  We have also added a grid of $45^\circ$ dotted lines that
shows all of the segments that the string could possibly follow.  In
Figure~\ref{fig.waves-rule}b we add in horizontal black bars, in order
to partition the space into a set of 2$\times$2 blocks that can be
updated independently.  Note that in all cases the segments that are
allowed to change during this update step, as well as the cells that
they will occupy after the update, are enclosed in a single block.
The long box below Figure~\ref{fig.waves-rule}b contains just the
slope information from the string.  This array of gradients is clearly
sufficient to recreate the wave pattern if the height at one position
is known.  This is not part of the 2D dynamics: it will be discussed
as a related 1D dynamics.

\figt{waves-rule}{%
\begin{array}{c@{\hspace{.3in}}c@{\hspace{.3in}}c}
\includegraphics[height=1.3in]{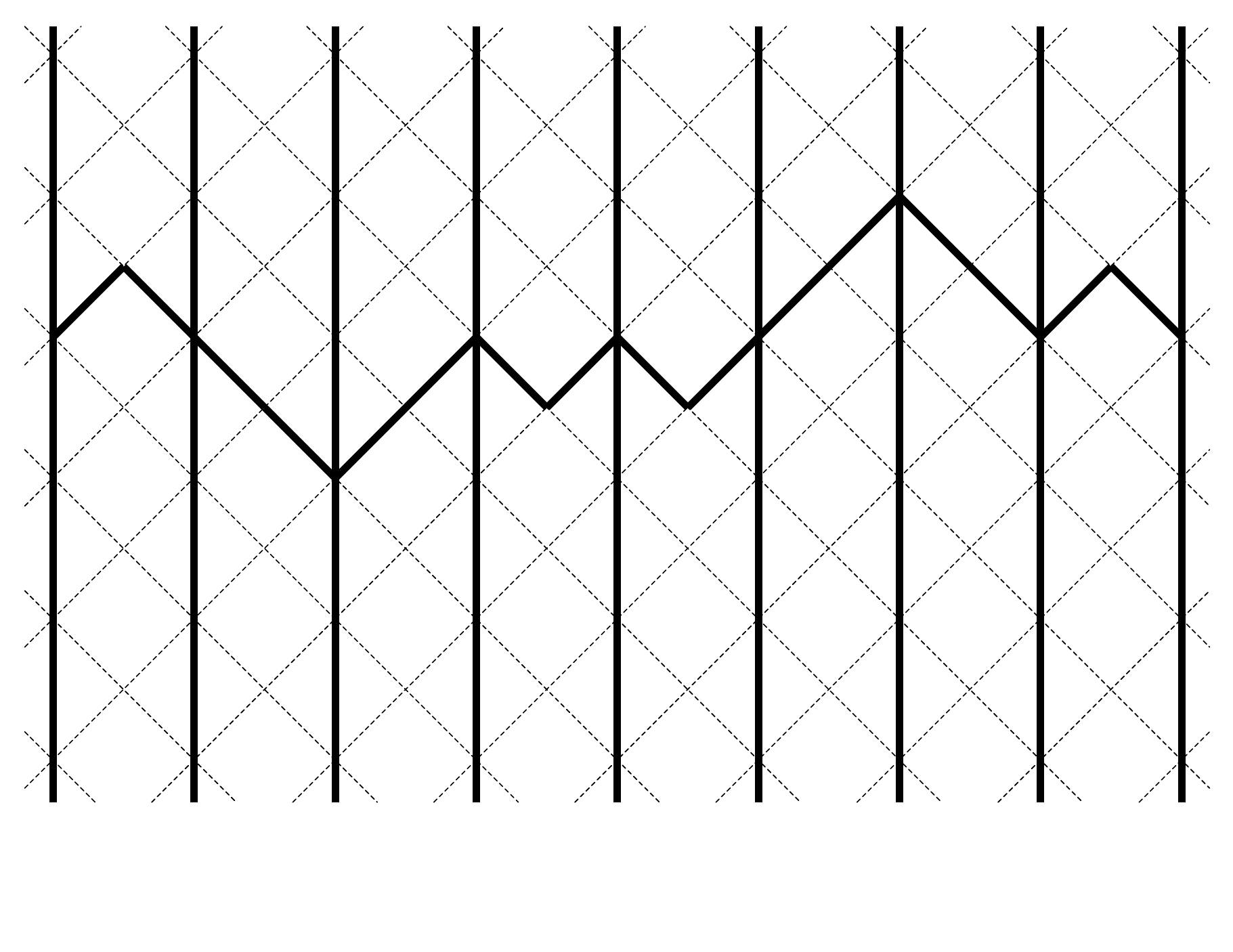} &
\includegraphics[height=1.3in]{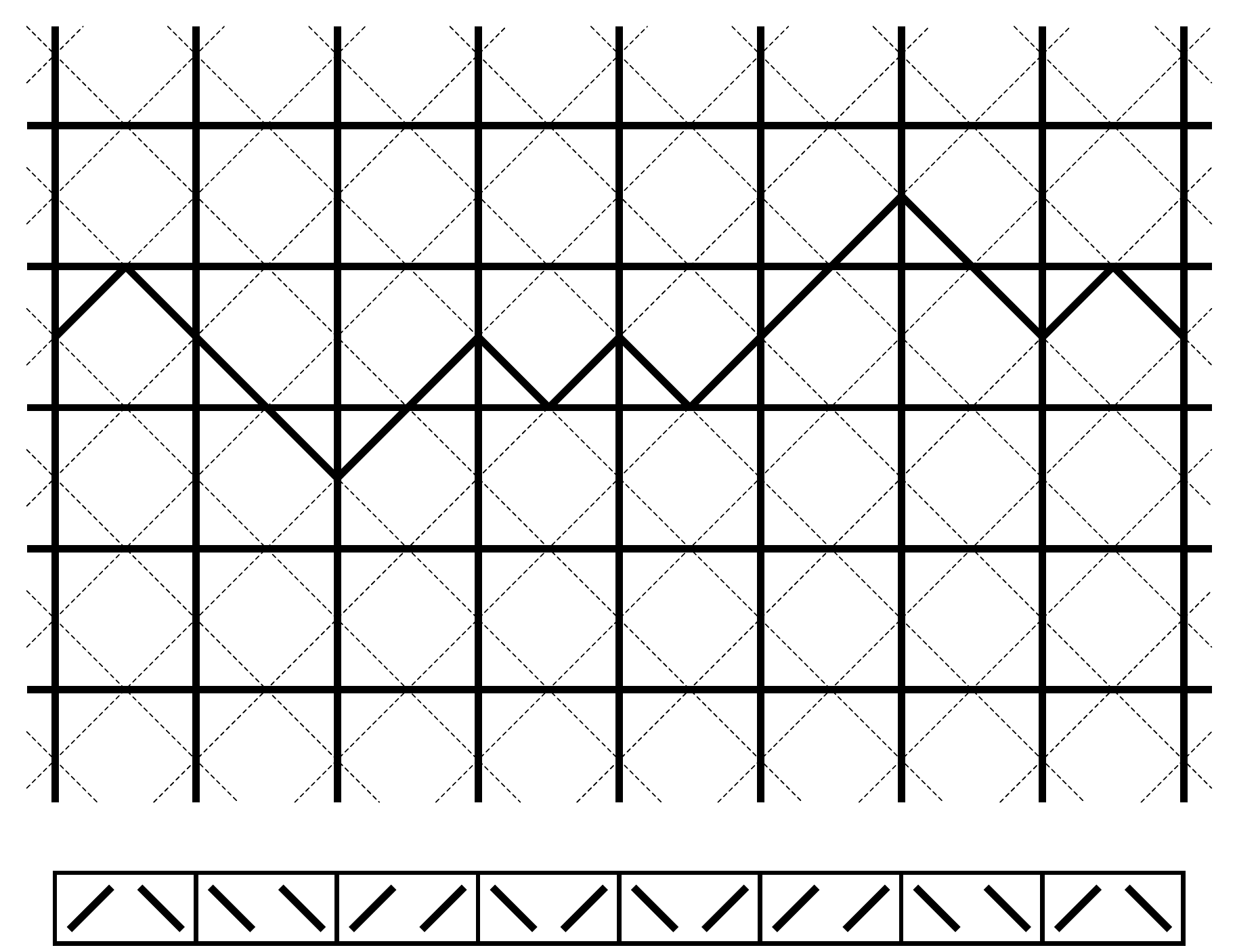} &
\includegraphics[height=1.3in]{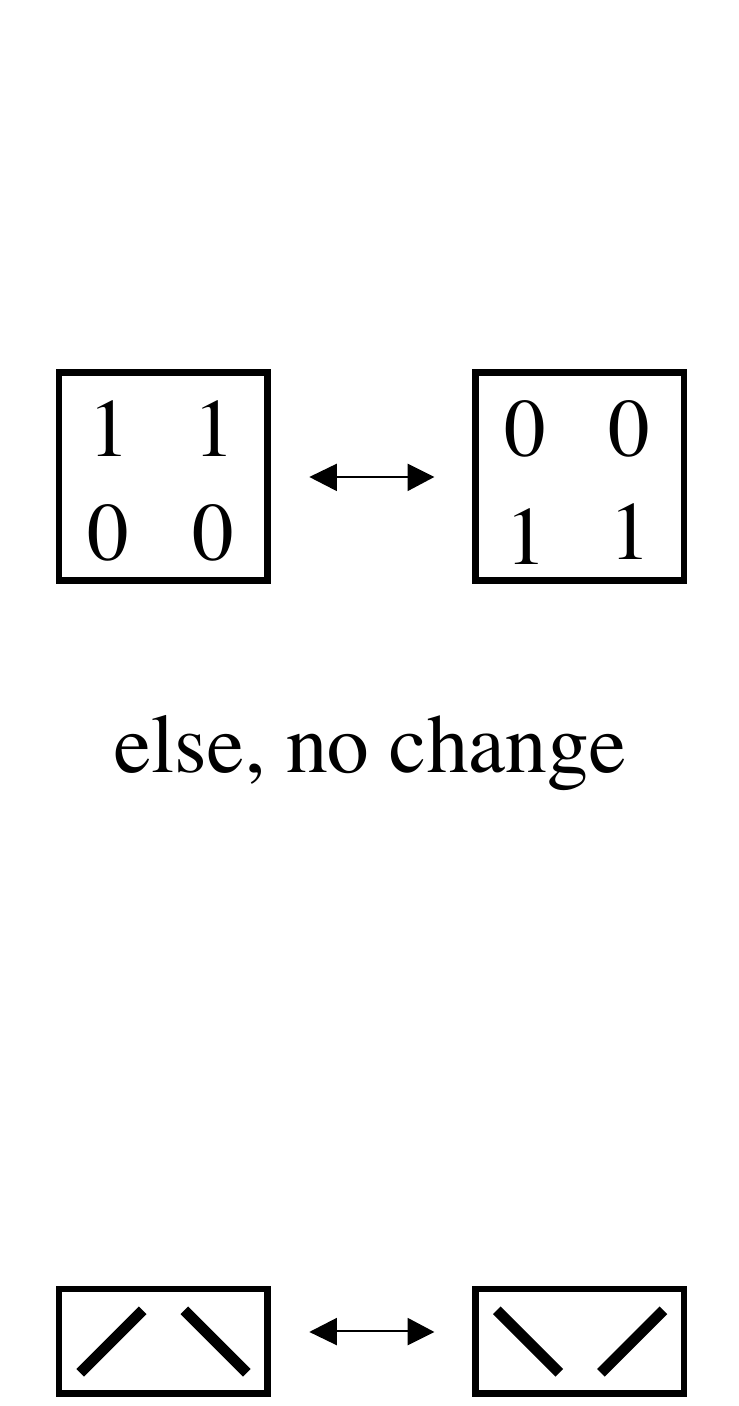} \\[.1in] \mbox{\bf (a)} &
\mbox{\bf (b)} & \mbox{\bf (c)} \\
\end{array}}
{Discrete wave dynamics.  (a) A wave configuration. Possible wave
  paths are indicated by dotted lines.  (b) Horizontal and vertical
  lines indicate one of two partitions used for discrete update rule.
  A 1D array summarizing wave gradients is shown below (not part of
  the 2D dynamics).  (c) Top, dynamical rule for 2D wave.  Presence of
  wave-path segments is indicated by 1's.  Bottom, equivalent 1D
  dynamical rule for gradients.}

Figure~\ref{fig.waves-rule}c shows the dynamical rule for a block.
Since the dotted lines indicate the direction in which segments must
run if they appear in any cell, the state information for each segment
is only whether it is there or not: this is indicated with a 1 or a 0.
The only segments that change are peaks \verb|/\| or valleys
\verb|\/|, and these are represented by two 1's at the top of a block
or at the bottom of a block respectively.  The rule is that peaks and
valleys turn into each other, and nothing else changes.  We apply the
rule alternately to the blocks shown, and to a complementary partition
shifted half a block horizontally and vertically.

\subsection{Exact Wave Behavior}

At the bottom of Figure~\ref{fig.waves-rule}c we've presented a
dynamics for the {\em gradients} of the wave.  The full 2D dynamics
just turns peaks into valleys and vice versa, leaving straight
segments unchanged: we can do that equally well on the array of
gradients.  As the 2D dynamics interchanges which blocking to use, the
dynamics on the gradients also alternates which pairs of gradients to
update together.  In all cases, the dynamics on the gradients
duplicates what happens on the string: if the two dynamics are both
performed in parallel, the gradient listed below a column will always
match the slope of the string in that column.

The dynamics on the gradients has an interesting property.  Turning a
peak into a valley and vice versa is exactly the same as swapping the
left and right elements of a block.  Leaving a \verb|//| or \verb|\\|
unchanged is also exactly the same as swapping the left and right
elements of a block.  In all cases, the dynamics on the gradients is
equivalent to a swap.

This means that the left element of a block will get swapped into the
right position, and at the next update it will be the left element of
a new block and will again get swapped into the right position, and so
on.  Thus all of the gradients that start off in the left side of a
block will travel uniformly to the right, and all that start in the
right side of a block will travel uniformly to the left.

This shows that the system obeys a discrete version of the wave
equation.  Half of the gradients constitute a right-going wave, and
half constitute a left-going wave.  At any step of the dynamics, the
2D wave in the original dynamics is just the sum of the two waves: it
is reproduced by laying gradients end to end.

If we refine the lattice, using more and more cells to represent a
wave of given width, smoother and smoother waves can be represented.
Of course even without going to a large-scale limit, the CA dynamics
is already exactly equivalent to a continuous wave equation with
constrained initial wave shapes, sampled at integer times: simply
stretch the rightgoing and leftgoing waves constructed out of
gradients to the full width of the lattice.  This just amounts to
drawing the wave shape corresponding to each block of the current
partition a little differently.

\subsection{Overall Transverse Motion}\label{sec.rel-string}

Assume the string carrying a discrete wave wraps around the space.
We've discussed the horizontal motion of waves along such a string,
but the string itself can move vertically.  For example, a pattern
such as \verb|\/\/\/...\/| all the way around the space reproduces
itself after two partition update steps, but shifted vertically by two
lattice units.  This is clearly the maximum rate of travel for a
string: one position vertically per update step.  Call this $v_0$.

\newcommand{\up}{{\texttt{\tiny $\setminus/$}}}
\newcommand{\dn}{{\texttt{\tiny $/\setminus$}}}

\newcommand{\p}{{\texttt{\tiny $+$}}}
\newcommand{\m}{{\texttt{\tiny $-$}}}

We can express the net velocity of the string in terms of the
populations of rightgoing and leftgoing gradient segments.  Let $R_\p$
be the number of rightgoing segments with slope $+1$ (rightgoing
$\verb|/|$'s), and similarly for $R_\m$, $L_\p$ and $L_\m$.  If the width
of the space is $B$ blocks, then there are $B=R_\p + R_\m$ segments
forming the rightgoing wave, and $B=L_\p + L_\m$ forming the leftgoing
one.

For the rightgoing or leftgoing wave, periodically repeating its
sequence of gradients corresponds to an unbounded wave with the same
average slope.  When both waves have shifted horizontally the width of
one period (after $2B$ partition update steps), the net vertical shift
is the sum of the slopes of the leftgoing gradients, minus the sum for
the rightgoing ones: $(L_\p - L_\m) - (R_\p - R_\m)$.  We can compute
this by summing the differences for each pair of slopes grouped
together in the columns of one partition.  Only columns containing
\verb|\/| or \verb|/\| contribute a non-zero difference, and so we
only need to count the numbers of blocks $B_\up$ and $B_\dn$ that are
about to change, to compute the constant velocity
\begin{equation}\label{eq.v}
{v\over v_0} = {(L_\p - L_\m) - (R_\p - R_\m) \over 2B} = {B_\up -
  B_\dn \over B}\,.
\end{equation}

\subsection{Ideal Energy and Momentum}

Only blocks that change have overall motion, and with $\tau$ the time
taken to update one partition, the frequencies of positive and
negative motion are $B_\up/\tau$ and $B_\dn/\tau$.  Thus from
\eqn{mu}, attributing a momentum to each changing block, the total
ideal momentum up is $h B_\up/2\tau v_0$, and down is $h B_\dn/2\tau
v_0$, so the net ideal momentum $p = (h/2\tau v_0)(B_\up-B_\dn)$.
From \eqn{v}, the corresponding relativistic energy is $E=c^2 p/v = (h
B/2\tau)/(v_0/c)^2$.  Letting $v_0 \to c$ to minimize energy, and
choosing units with $h=2$ and $c=1$ and $\tau=1$, this becomes
\begin{equation}\label{eq.ep}
E=B  \quad \textrm{and} \quad  p= B_\up - B_\dn\,.
\end{equation}
Energy is the constant width (in blocks) of the string, and momentum
is the constant net number of blocks moving up.

There is an interesting subtlety involved in letting $v_0\to c$ in the
2D dynamics.  We interpret all gradient segments as always moving,
swapping in pairs in each update in order to recover the wave
equation---even though some paired segments are in different blocks
when they ``swap'' identical values. If all block motion forward or
backward is at the speed $c$, each segment must be interpreted as
traveling at the speed $c\sqrt{2}$ as it swaps diagonally.  If instead
we interpret segments as moving up and down (or not moving), none
travel faster than light, but the interaction is non-local at the
scale of an individual block.

\subsection{Rest Frame Energy}

For the transverse motion of the string to approach the maximum speed,
almost all of the block updates must contribute to overall motion, and
almost none to just internally changing the string.  This slowdown of
internal dynamics is a kind of time dilation, which is reflected in
the rest frame energy $\sqrt{E^2 - p^2}$.  From \eqn{ep},
\begin{equation}\label{eq.erb}
E_r=\sqrt{B^2 - (B_\up-B_\dn)^2}\,.
\end{equation}
The energy $E_r$ available for rest-frame state-change decreases as
more blocks move in the same direction.  In this model total energy
$E$ is independent of $v$, hence rest energy $E_r= E/\gamma$ must
approach 0 as $1/\gamma \to 0$.  This contrasts with a normal
relativistic system that can never attain the speed of light, which
has a constant rest energy $E_r$ and a total energy $E$ that changes
with $v$.

The analysis up to here applies equally well to both the 1D and 2D
versions of the dynamics of Fig.~\ref{fig.waves-rule}.  In 2D,
however, there is an additional constraint: there must be an equal
number $B$ of positive and negative slopes, so that the string meets
itself at the periodic boundary.  Since there are also an equal number
$B$ of right and left going gradients, $R_\p=L_\m$ and $R_-=L_\p$.
Thus from \eqn{v} and \eqn{erb},
\begin{equation}\label{eq.er}
E_r=2\sqrt{R_\p L_\p}\,.
\end{equation}
If $R_\p/B$ were the probability for a walker to take a step to the
right, and $L_\p/B$ the probability to the left, then \eqn{er} would
be the standard deviation for a $2B$-step random walk.  Related models
of diffusive behavior that make contact with relativity are discussed
in ~\cite{smith,curious-properties,toffoli-action}.  None of these
define relativistic objects that have an internal dynamics, however.

\section{Discussion}

Given the definition of a finite-state dynamics, we could try to
assign intrinsic properties to it based on the best possible
implementation.  For example, programming it on an ordinary computer,
a basic property is the minimum time needed, on average, to simulate a
step of the dynamics.  It would be hard, though, to be sure we've
found the most efficient mapping onto the computer's architecture, and
the minimum time would change if we used a different computer, or
built custom hardware using various technologies.  The true minimum
time would correspond to the fastest possible implementation allowed
by nature!  Such a definition seems vacuous, though, since we don't
know the ultimate laws of nature, and even if we did, how would we
find the best possible way to use them?

Surprisingly, a fundamental-physics based definition of intrinsic
properties is not in fact vacuous, if we base it on general
principles.  Assuming the universe is fundamentally quantum
mechanical, we couldn't do better than to simply {\em define} a
hamiltonian that exactly implements the classical finite-state
dynamics desired at discrete times, with no extra distinct states or
distinct state change.  This ideal hamiltonian identifies the fastest
implementation that is {\em mathematically possible}, with given
average energy.

This procedure assigns to every invertible finite-state dynamics an
ideal energy that depends only on the average rate of distinct state
change.  This is generally not much like a physical energy, though,
since we haven't yet included any realistic constraints on the
dynamics.  For example, each state change might correspond to a
complete update of an entire spatial lattice, as in the synchronous
definition of a CA.  Then the energy would be independent of the size
of the system.  We can fix this by constraining the finite-state
dynamics to be local and not {\em require} synchrony: defining it in
terms of gates that are applied independently.

We expect the ideal energy, and distinct portions of it, to become
more realistic with additional realistic constraints.  For this
reason, we studied invertible lattice dynamics derived from the
integer-time behavior of idealized classical mechanical systems.  In
the examples we looked at, ideal energies and momenta defined by local
rates of state change evolve like classical relativistic quantities.

It seems interesting and novel to introduce intrinsic definitions of
energy and other physical quantities into classical finite-state
systems, and to use these definitions in constructing and analyzing
finite-state models of physical dynamics.  Since all finite-energy
systems in the classical world actually have finite state, and since
classical mechanics doesn't, this may be a productive line of inquiry
for better modeling and understanding that world.  Moreover, inasmuch
as all physical dynamics can be regarded as finite-dimensional quantum
computation, finite-state models of classical mechanics may play the
role of ordinary computation in understanding the more general quantum
case.

\subsubsection*{Acknowledgments.}  I thank Micah Brodsky and Gerald
Sussman for helpful discussions.


\begin{thebibliography}{99}
\bibitem{bekenstein} Bekenstein, J. D.: Universal upper bound on the
  entropy-to-energy ratio for bounded systems. Phys. Rev. D {\bf 23},
  287 (1981).
\bibitem{max-speed} Margolus, N., Levitin, L. B.: The maximum speed of
  dynamical evolution. Physica D\ {\bf 120}, 188 (1998).
\bibitem{max-avg} Margolus, N.: The maximum average rate of state
  change.  \eprint{arXiv:1109.4994}
\bibitem{planck} Planck, M.: On the law of distribution of energy in
  the normal spectrum.  Ann.\ Phys.\ (Berlin) {\bf 309}, 553
  (1901).
\bibitem{emulation} Margolus, N.: Quantum emulation of classical
  dynamics.  \eprint{arXiv:1109.4995}
\bibitem{phys-like} Margolus, N.: Physics like models of
  computation.  Physica D\ {\bf 10}, 81 (1984).
\bibitem{cc} Margolus, N.: Crystalline Computation. In: Hey, A. (ed.)
  {\em Feynman and Computation}. Perseus Books, 267 (1998).
  \eprint{arXiv:comp-gas/9811002}
\bibitem{cam-book} Toffoli, T., Margolus,  N.: {\em Cellular automata
  machines: a new environment for modeling}. MIT Press (1987).
\bibitem{chopard-book} Chopard, B., Droz, M.: {\em Cellular Automata
  Modeling of Physical Systems}.
  Cambridge University Press, (2005).
\bibitem{rothman-book} Rothman, D., Zaleski, S.: {\em Lattice Gas
  Cellular Automata: Simple Models of Complex Hydrodynamics}.
  Cambridge University Press, (2004).
\bibitem{rivet-book} Rivet, J. P.,  Boon, J. P.: {\em Lattice Gas
  Hydrodynamics}.  Cambridge U. Press, (2005).
\bibitem{salt} Fredkin, E.: A computing architecture for physics.
  In: {\em CF '05 Proceedings of the 2nd conference on computing
    frontiers}. ACM, 273 (2005).
\bibitem{nks} Wolfram, S.: {\em A new kind of science}. Wolfram Media, (2002).
\bibitem{ica} Toffoli, T., Margolus, N.: Invertible cellular automata:
  a review.  Physica D {\bf 45}, 229 (1990).
\bibitem{kari} Kari, J.: Representation of reversible cellular
  automata with block permutations.  Math. Systems Theory {\bf 29},
  47 (1996).
\bibitem{lose} Durand-Lose, J.: Representing reversible cellular
  automata with reversible block cellular automata.  Discrete
  Math. Theor. Comp Sci. Proc. AA, 145 (2001).
\bibitem{ulam} Ulam, S.: Random Processes and Transformations. In:
  Proceedings of the International Congress on Mathematics, 1950,
  Vol. 2, 264 (1952).
\bibitem{von-neumann} von Neumann, J.: {\em Theory of Self-Reproducing
  Automata}, University of Illinois Press (1966).
\bibitem{zuse} Zuse, K.: Calculating Space. MIT
  Tech. Transl. AZT-70-164-GEMIT (1970).
\bibitem{bcq} Margolus, N.: Mechanical Systems that are both Classical
  and Quantum. \eprint{arXiv:0805.3357}
\bibitem{ssm} Margolus, N.: Universal cellular automata based on the
  collisions of soft spheres. In: Griffeath, D., Moore, C. (eds.)
  {\em New Constructions in Cellular Automata}.  Oxford University
  Press, 231 (2003). \eprint{arXiv:0806.0127}
\bibitem{bbm} Fredkin, E., Toffoli, T.: Conservative logic.
  Int. J. Theor. Phys. {\bf 21}, 219 (1982).
\bibitem{hh} Hrgov\v{c}i\'{c}, H.: Discrete representations of the
  n-dimensional wave equation.  J.\ Phys.\ A: Math.\ Gen.\ {\bf
    25}, 1329 (1992).
\bibitem{toffoli-action} Toffoli, T.: Action, or the fungibility of
  computation.  In: Hey, A (ed) {\em Feynman and computation}.  Perseus Books, 349 (1998).
\bibitem{marg-thesis} Margolus, N.: {\em Physics and computation}.
  Massachusetts Institute of Technology Ph.D. Thesis (1987).
\bibitem{toffoli-pde} Toffoli, T.: Cellular automata as
an alternative to (rather than an approximation of) differential
equations in modeling physics.  Physica D\ {\bf 10}, 117 (1984). 
\bibitem{smith} Smith, M.: Representation of geometrical and
  topological quantities in cellular automata.  Physica D {\bf 45},
  271 (1990).
\bibitem{curious-properties} Ben-Abraham, S. I.: Curious properties of
  simple random walks.  J. Stat.\ Phys.\ {\bf 73}, 441 (1993).
  \end{thebibliography}
\end{document}